\documentclass[journal]{IEEEtran}

\IEEEoverridecommandlockouts

\usepackage{cite,fancyhdr}
\usepackage{amsmath,amssymb,amsfonts}
\usepackage{algorithmic}
\usepackage{graphicx}
\usepackage{textcomp}
\usepackage{xcolor}
\def\BibTeX{{\rm B\kern-.05em{\sc i\kern-.025em b}\kern-.08em
    T\kern-.1667em\lower.7ex\hbox{E}\kern-.125emX}}

\newcommand{\equals}{\!=\!}

\begin{document}

\title{Confidence-Driven Adaptive Semantic Transmission for Task-Oriented Communication 

\thanks{
This publication is an outcome of the R\&D work undertaken during the tenure of PDF award under the Visvesvaraya PhD Scheme, being implemented by PhD Cell, Digital India Corporation, MeitY.}
}

\author{Suman Das, Prajnyajit Mohanty, and Siddhartha S. Borkotoky \\ \textit{Indian Institute of Technology Bhubaneswar, Khordha 752051, Odisha, India}\\
\texttt{(25sp06010, prajnyajit, borkotoky)@iitbbs.ac.in}  
}

\maketitle
\thispagestyle{fancy}
\lhead{\textit{Accepted for Presentation at IEEE ANTS 2026.}}

\begin{abstract}
Image-based remote monitoring over wireless networks faces a trade-off between communication cost and classification reliability. Transmitting complete images is communication-intensive, whereas relying solely on lightweight edge classifiers may compromise accuracy. To address this challenge, we propose Confidence-Driven Adaptation (CDA) for image transmission and classification. An edge device (ED) captures images and applies a lightweight classifier before transmission to a remote server. Based on its classification confidence, the ED selects among three transmission modes with varying levels of information: a short text string reporting its predicted class when confidence is high, a compressed Region of Interest (ROI) at intermediate confidence levels, or an uncompressed ROI when confidence is low. The server accepts the predicted class directly in the first case and performs classification using a more powerful classifier in the other two cases. Numerical evaluations under bandwidth-limited, lossy links and latency-constrained settings demonstrate that CDA achieves high task accuracy at the receiver compared with non-adaptive baselines that transmit only the edge-classifier outcome or the uncompressed ROI. \end{abstract}

% \vspace{-2mm}
\vspace{-2mm}
\section{Introduction}
Task-oriented semantic communication is an emerging paradigm in which a transmitter sends only the task-relevant information necessary to execute a specific action~\cite{b5}. By shifting the focus away from bit-accurate delivery of complete source data, task-oriented approaches can substantially reduce communication resource consumption. Consequently, they have generated significant interest in recent years.
Specifically, task-oriented communication for image-based monitoring has received considerable attention. Existing approaches for such applications typically focus on extracting and transmitting the region of interest (ROI) from the source image~\cite{b5},~\cite{b4}. Since the ROI is usually much smaller than the full image, its transmission reduces bandwidth consumption and latency. However, a typical ROI is often several kilobytes in size. Consequently, repeated ROI transmissions can incur substantial latency when the wireless link has limited bandwidth and experiences frequent packet erasures, as is common in communication-constrained remote deployments such as wildlife, agricultural, and structural health monitoring. An alternative is to classify the image at the source and transmit only the classifier output. However, the performance of this approach is constrained by the accuracy of the source's classifier. This imposes a limitation for applications in which the image sources are resource-constrained edge devices (EDs) that cannot run sophisticated and compute-heavy classifiers. 

We address this gap by proposing Confidence-Driven Adaptation (CDA) for image-classification tasks over wireless networks. CDA reduces communication costs from the image source (an ED) to a server while preserving task accuracy. Instead of transmitting an ROI for every captured image, the ED dynamically selects from a compact text label representing the output of a local lightweight classifier, a compressed ROI with a compression level chosen from a predefined set, and the full ROI. The selection is based on the classifier's confidence in its prediction; the higher the confidence, the more compact the representation sent. 

Existing task-oriented semantic communication schemes \cite{b6}, \cite{b7} rely on jointly trained neural encoder-decoders and focus on reconstruction-oriented objectives rather than decision correctness. In contrast, our lightweight approach directly targets receiver-side decision accuracy. Although classifier confidence is considered in \cite{b8}, it supports only a binary send-or-not decision, whereas our adaptive policy uses confidence to control the amount of transmitted information.

\begin{figure}
    \centering
    \includegraphics[width=0.85\linewidth]{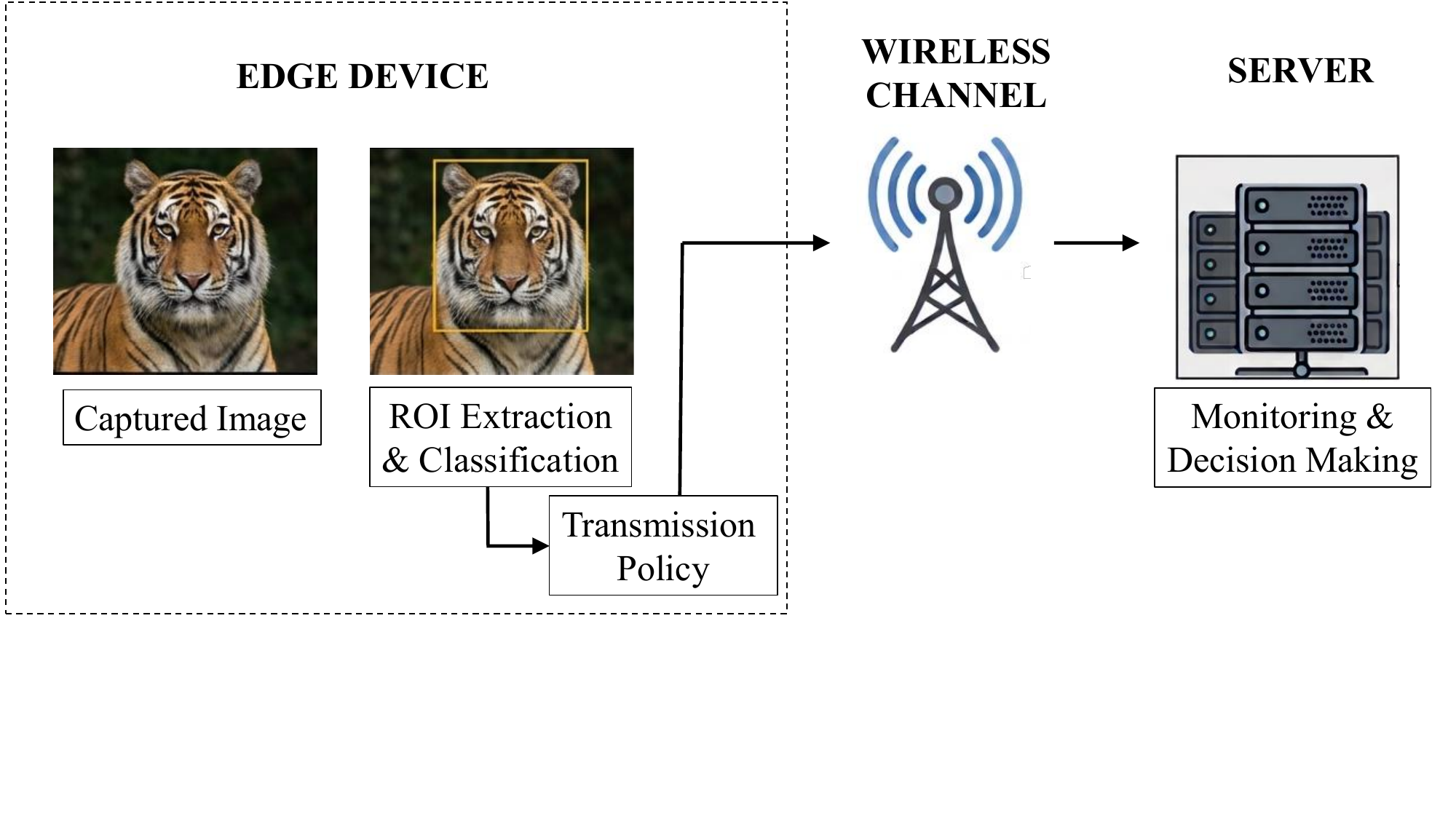}
    \vspace{-12mm}
    \caption{Overview of an image-based monitoring system.}
    \label{fig:overview}
\end{figure}

\section{System Model}
Fig.~\ref{fig:overview} provides an overview of the system. We consider a monitoring application in which an ED captures an image. The objective is to ensure correct knowledge at a remote server about the class to which the object in the image belongs. There are $K$ object classes belonging to the set $\mathcal{C}=\{1,\ldots,K\}$.  

The ED contains an ROI detector and a lightweight classifier. For a captured image~$I$, the detector yields the ROI $x_0=\mathcal{D}(I)$. Its ground-truth class is $y\in\mathcal{C}$. For our illustration, we implement an ImageNet-pretrained MobileNetV2~\cite{b1} on $224 \!\times\! 224 \!\times\! 3$ RGB inputs, followed by global average pooling, dropout, and a $K$-class softmax layer. The classification head is trained using Adam (learning rate 0.001) for up to 25 epochs, with early stopping (patience = 4).
A COCO-pretrained YOLOv8n detector identifies the most confident ROI regardless of class, with the full image serving as the ROI when no sufficiently confident and large detection is found. For input image $x$, the classifier produces the posterior probabilities $\mathbf{p} = f(x) = (p_1, \dots, p_K)$,
% \begin{equation}
% \mathbf{p} = f(x) = (p_1, \dots, p_K), \qquad p_k \ge 0, \quad \sum_{k=1}^{K} p_k = 1 ,
% \end{equation}
where $p_k$ is the posterior probability of the $k$-th class. The predicted class is the one with the maximum posterior probability, that is,
\begin{equation}
\hat{y}(x) = \arg\max_{k \in \mathcal{C}} p_k.
\end{equation}
The classifier's confidence in its prediction is the corresponding maximum posterior probability, that is,
\begin{equation}
c(x) = p_{\hat{y}(x)}.
\label{eq:conf}
\end{equation}

At the server, the DINOv2 ViT-B/14 backbone~\cite{dinov2} with L2 normalization and a linear classification head is used to process images of size $224 \times 224 \times 3$. The linear classifier is trained for 30 epochs using Adam (learning rate 0.001), retaining the best validation checkpoint.

Note that the ED and the server exhibit asymmetric computational complexity. The model at the ED contains approximately $3.4$M \cite{b1} parameters, whereas the server’s DINOv2 backbone contains approximately $86$M \cite{receiver-model} parameters.  

The ED communicates with the server over a wireless link. Each packet on the link can carry up to $b_{\max}$ bytes of application data. The payload is encoded using a \mbox{rate-$r_c$} channel code for forward erasure correction. Each packet includes a \mbox{$b_h$-byte} header. The packet duration in seconds is given by $L_p = (8b_h + \lceil 8b/r_c \rceil)/R_b$, where $b$ is the size of the application data (in bytes) in the packet, and $R_b$ is the physical-layer transmission rate in bits per second. We model an independent Bernoulli packet-erasure channel. The ED continues transmitting until either all source data have been delivered (indicated by a receiver feedback), or the transmission latency (computed as the sum of the airtimes of the packets transmitted) exceeds a threshold $T_{\max}$, whichever occurs earlier.  The value of $T_{\max}$ depends on the latency tolerance of the monitoring application.

\section{Proposed Scheme}
CDA comprises three components, namely (a) a parametric compression operator mapping an image to a discrete ladder of size or quality levels, (b) an offline calibration step learning a mapping from confidence to a byte budget, and (c) a confidence-driven adaptive transmission policy.

\subsection{Compression Operator and Level Ladder}
We define a parametric compression operator $\mathcal{T}_{\ell}$ indexed by a compression level $\ell=(r,q)$, where $r$ denotes the spatial resolution (in pixels) and $q$ the JPEG quality factor. Applying $\mathcal{T}_{\ell}$ to an image $x$ involves:
(i) resizing $x$ to $r\times r$ pixels using bilinear interpolation
and (ii) encoding the resized image as a JPEG at quality $q$. Specifically, $\mathcal{T}_{\ell}$ represents the pair
\begin{equation}
\mathcal{T}_{\ell}(x) \mapsto \big(x_{\ell},s_{\ell}(x)\big),
\end{equation}
where $x_{\ell}$ is the compressed image and $s_{\ell}(x)$ is its size in bytes. Therefore, $s_{\ell}(x)$ represents the communication payload if $x_{\ell}$ were transmitted. The operator is instantiated over a fixed, monotonically decreasing ladder of
$M=10$ compression levels, $\mathcal{L}=\{\ell_1,\ldots,\ell_M\}$, ordered from mildest to most aggressive compression:
\begin{equation}
\begin{aligned}
\mathcal{L} = \big\{&
(224,95),(192,90),(160,85),(128,80),(112,75),\\
& (96,65),(80,55),(64,45),(48,35),(32,25)
\big\}.
\end{aligned}
\end{equation}

\subsection{Offline Calibration of the Confidence-to-Bytes Mapping}
The communication policy uses an offline-calibrated mapping that determines the minimum required payload to be sent for a given classifier confidence. Calibration is performed once using a dedicated subset of the held-out set with known ground-truth labels. For each calibration image, our method evaluates all levels of the compression
ladder and records the smallest payload $g(x)$  (in bytes) for which the classifier remains
correct. That is,
\begin{equation}
g(x)=\min_{\ell\in\mathcal{L}} s_{\ell}(x)
\quad \text{s.t.}\quad
\hat{y}(x_{\ell})=y, 
\label{eq:minsafe}
\end{equation}
where $\hat{y}(x_{\ell})$ is the predicted class obtained by feeding the compressed ROI $x_{\ell}$ to the predictor and $y$ is the ground-truth label. If no compressed version is correctly classified, then $g(x)$ is taken to be the full ROI size. Pairing the resulting $g(x)$ with the corresponding confidence $c(x)$ yields a $(\mathrm{bytes},\mathrm{confidence})$ pair. For each image $x$ in the held-out set, such a pair is obtained. Next, using isotonic regression on the resulting pairs, we fit a confidence-to-byte mapping $\phi(c)$ such that higher confidence corresponds to a smaller byte budget, while lower confidence allows a larger budget. Note that the calibrated mapping $\phi(c)$ is the only data-driven component of our proposed policy. All other parameters are predefined.

\subsection{Confidence-Driven Adaptive (CDA) Transmission}
During deployment, the ED selects the transmission mode using only its local prediction confidence $c=c(x_0)$, without access to the ground-truth label. The transmission mode is defined using two thresholds, as follows: 
\begin{equation}
\mathrm{mode}(c)=
\begin{cases}
\mathrm{full\, ROI}, & c<\tau_{\mathrm{low}},\\
\mathrm{compressed\, ROI}, & \tau_{\mathrm{low}}\leq c\leq\tau_{\mathrm{high}},\\
\mathrm{text}, & c>\tau_{\mathrm{high}}.
\end{cases}
\end{equation}

That is, for low confidence, the node transmits the full-quality ROI to preserve
discriminative information at the receiver. By contrast, it transmits only the predicted label as a compact text representation when confidence is high. For our numerical results, we assume $\tau_{\mathrm{low}} \!=\! 0.3$ and $\tau_{\mathrm{high}} \!=\! 0.9$. Fine-tuning these values is left as future work.

For intermediate confidence, the ED obtains the byte budget $\phi(c)$ from the calibrated mapping, which provides an estimated byte budget based on the locally observed classifier confidence. Next, it selects the least-compressed ROI whose size fits within this budget. 

Upon receiving an ROI, the server applies its classifier to predict the image class $\hat{y}_{\mathrm{rx}}$. For text
reception, the receiver directly adopts the transmitted label, yielding
$\hat{y}_{\mathrm{rx}}=\hat{y}(x_0)$.

\section{Performance Evaluation}
We consider the task of classifying an animal image into one of four classes, namely $\mathcal{C} = \{ \mathrm{Lion, Deer, Elephant, Giraffe}\}$. 
The dataset comprises 1321 images distributed across the four classes \cite{dataset}. At the transmitter, 581 images are used for training, 102 for validation, and 160 are reserved as a held-out set for calibrating the confidence-to-byte mapping. The receiver is independently trained using the same dataset, with 928 images allocated for training and the remaining 233 for validation. For the wireless link, we assume $R_b \equals 1$~kbps, $b_h \equals 4$ bytes, $b_{\max} \equals 250$ bytes, and $r_c \equals 0.5$. These values are chosen as representative of low bit-rate, long range links (such as low-power wide area networks) deployed over remote regions without reliable cellular infrastructure (e.g., a wildlife-monitoring scenario).

% \begin{equation}
% \mathrm{Accuracy}_{\mathrm{rx}} =
% \frac{1}{|\mathcal{H}|}
% \sum_{x\in\mathcal{H}}
% \mathbf{1}\!\left[\hat{y}_{\mathrm{rx}}(x)=y\right],
% \end{equation}
% where $\mathcal{H}$ denotes the evaluation set and the
% indicator function $\mathbf{1}[\cdot]$ is $1$ for a correct decision and $0$
% otherwise. 

\begin{figure}
    \centering
    \includegraphics[width=0.93\linewidth]{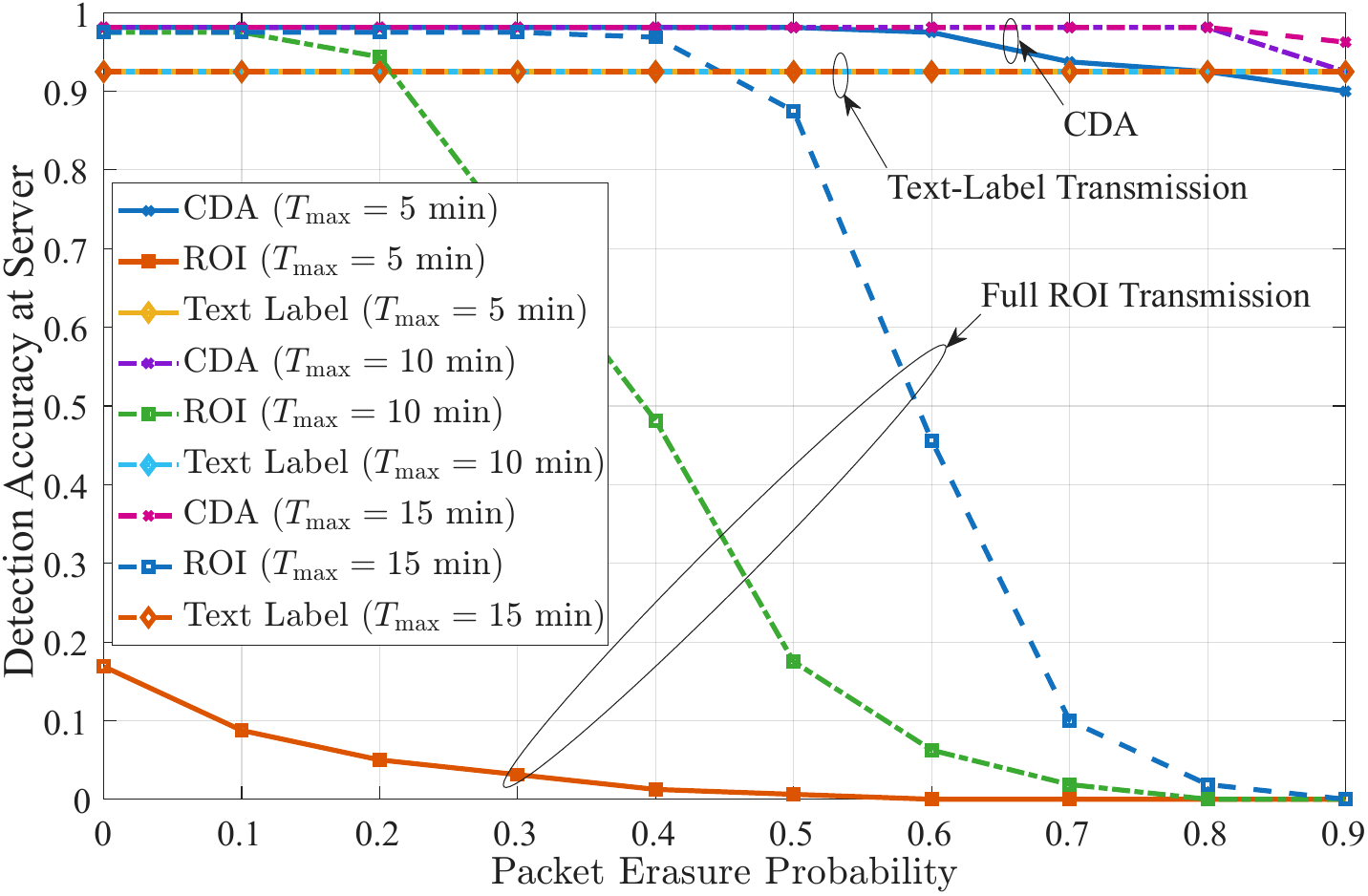}
    \vspace{-2mm}
    \caption{Comparison of classification accuracy at the server.}
    \label{fig:RxAccuracy_vs_PER}
    % \vspace{-2mm}
\end{figure}

Fig.~\ref{fig:RxAccuracy_vs_PER} compares CDA's detection accuracy at the receiver (i.e.,
the fraction of correct classification instances over the evaluation set) with two baselines. One baseline transmits the full ROI, with classification performed at the receiver. In the other, classification is performed only at the transmitter, and the classifier output is transmitted as a text label. We consider latency deadlines $T_{\max} \equals$ 5, 10, and 15 minutes. 

We observe that full-ROI transmission is highly sensitive to the PER and the latency deadline $T_{\max}$. For large $T_{\max}$ and small PER, the scheme achieves high accuracy ($\approx$ 98\%). Under these conditions, the transmitter is able to deliver the complete ROI to the receiver within the deadline. The richness of information contained in the ROI leads to high classification accuracy.
However, the performance of full-ROI transmission is very poor in the high-PER regime. This occurs because retransmissions are required to compensate for packet losses, forcing the sender to frequently violate the latency deadline. For $T_{\max} \equals $ 5 minutes, most of the ROIs could not be delivered even over a lossless channel (i.e.,  PER $=$ 0), leading to poor performance throughout. 
By contrast,  the text-label-only baseline provides a constant accuracy of approximately 92\% regardless of the PER and the latency deadline. Since transmitting a text label incurs negligible airtime, the label is delivered within the deadline even with many retransmissions in high-PER conditions. Thus, the accuracy observed at the server is identical to the ED's classifier accuracy ($\approx$ 92\%).

Unlike the baselines, the proposed CDA scheme provides high task accuracy for a wide range of PER values. Even with the tightest latency deadline of $T_{\max}\equals$ 5 minutes, CDA provides 98\% accuracy until PER reaches 0.6, and exhibits graceful degradation beyond this point. For larger values of $T_{\max}$, 98\% accuracy is sustained up to a PER of about 0.8. These results demonstrate the efficacy of CDA's confidence-driven adaptive approach.

\section{Conclusion}
We proposed a Confidence-Driven Adaptive (CDA) semantic transmission scheme for task-oriented image communication under constrained wireless links. CDA adapts the transmitted representation to the confidence of the source-side classifier, using a compact text label for high-confidence predictions, a compressed ROI for intermediate confidence, and the full ROI for low-confidence predictions.
Our preliminary investigation involving a wildlife image-classification task shows that CDA substantially reduces transmission latency while maintaining high receiver-side accuracy even under high packet erasure rates. Future work includes the investigation of channel-aware parameter selection and evaluation over a wider range of applications and wireless networks.

\end{document}